\input{aipcheck}

\documentclass[
    ,final            
  ]
  {aipproc}

\layoutstyle{6x9}

\newcommand{\be}{\begin{equation}}
\newcommand{\ee}{\end{equation}}
\newcommand{\bea}{\begin{eqnarray}}
\newcommand{\eea}{\end{eqnarray}}

\def\lsim{\mathrel{\rlap{\lower4pt\hbox{\hskip1pt$\sim$}}\raise1pt\hbox{$<$}}}
\def\gsim{\mathrel{\rlap{\lower4pt\hbox{\hskip1pt$\sim$}}\raise1pt\hbox{$>$}}}
\def\nostrocostruttino#1\over#2{\mathrel{\mathop{\kern 0pt \rlap
{\hbox{$#1$}}} \hbox{\kern-.135em $#2$}}}

\def\kt{k_\perp}

\def\xb{x_{_{\!B}}}

\def\avk{\langle k_\perp ^2\rangle}

\def\kt{k_\perp}
\def\ktm2{k^2_{\perp{\rm max}}}

\newcommand{\ptsmG}{\langle P_T^2\rangle_G}
\def\xb{x_{_{\!B}}}

\begin{document}

\title{Partonic Transverse Motion in Unpolarized\\ Semi-Inclusive Deep Inelastic Scattering}


\author{M. Boglione}{
  address={Dipartimento di Fisica Teorica, Universit\`a di Torino,
              Via P.~Giuria 1, I-10125 Torino, Italy\\
              INFN, Sezione di Torino, Via P.~Giuria 1, I-10125 Torino, Italy}
}

\author{S. Melis}{
  address={European Centre for Theoretical Studies in Nuclear Physics and
Related Areas (ECT*) \\
              Villa Tambosi, Strada delle Tabarelle 286, I-38123 Villazzano,
Trento, Italy}
}

\author{A. Prokudin}{
  address={Jefferson Laboratory, 12000 Jefferson Avenue, Newport News, VA 23606}
}

\begin{abstract}
We analyse the role of partonic transverse motion
in unpolarized Semi-Inclusive Deep Inelastic Scattering (SIDIS) processes.
Imposing appropriate kinematical conditions,
we find some constraints which fix an upper limit to the range of allowed $\kt$ 
values and lead to interesting results, particularly for some observables like 
the $\langle \cos \phi_h \rangle$ azimuthal modulation of the unpolarized SIDIS 
cross section and the average transverse momentum of the final, detected hadron.
\end{abstract}

\maketitle


Azimuthal spin asymmetries in SIDIS processes are directly
related to Transverse Momentum Dependent (TMD) parton distribution and
fragmentation functions,
and are the subject of intense theoretical and experimental studies. The usual,
collinear parton
distribution functions depend on the fraction $\xb$ of hadron momentum carried
by the scattering parton and
on the virtuality of the probe, $Q^2$. Additionally, TMDs depend on the intrinsic
transverse momentum of
the parton, \textit{\textbf k$_\perp$}, opening invaluable opportunities to unravel the
three-dimensional partonic
picture of the nucleon in momentum space.

In phenomenological analysis, the transverse momentum distribution
of the TMDs is usually assumed to be a Gaussian.
This is a convenient approximation as it allows to solve the $\kt$ integration analytically, and it leads to a successful description of many sets of data.
Inspired by the parton model, 
we bound the integration range of transverse momenta $k_{\perp}$ and
we observe, in some kinematical regions, remarkable deviations
from the predictions obtained from the common TMD approach, based on the
Gaussian parametrization integrated over the full $\kt$ range, $[0,\infty]$.
We show that some kinematical ranges, typically
low $\xb$ or equivalently low $Q^2$ regions, are not safely controlled
by the present phenomenological model, while a physical upper limit on the 
$k_{\perp}$ range can prevent uncontrolled large $k_{\perp}/Q$
contributions.
This leads, for instance, to a better description of some observables like the
$\langle\cos\phi_h\rangle$ asymmetry and introduces
some interesting effects in the $\langle P_T^2\rangle$ behaviors.

We study the SIDIS process in the $\gamma^*$ - proton c.m. frame,
where $\gamma^*$ denotes the virtual photon. The detailed kinematics 
is given in Refs.~\cite{Anselmino:2005sh,Anselmino:2011ch,Boglione:2011wm}.

A physical picture that allows us to put some further constraints on the partonic intrinsic 
motion is provided by the parton model, where kinematical limits on the transverse momentum size 
can be obtained by requiring the
energy of the parton to be less than the energy of the parent hadron and by
preventing the parton to move backward with respect to the parent hadron direction ($k_z < 0$).
They give, respectively:
\begin{equation}
k_{\perp}^2\le(2-\xb)(1-\xb)Q^2 \,\, \,\, , \, \,\, { 0 < \xb  <  1}\,.
\label{cutenergy}
\end{equation}
and
\begin{equation}
\kt ^2\le \frac{\xb(1-\xb)}{(1-2\xb)^2}Q^2 \, \, , \, \, { \xb  <  0.5}\,.
\label{cutdirection}
\end{equation}
Notice that these are exact relations, which hold at all orders in $(\kt/Q)$.
%

\begin{figure}
\includegraphics[height=0.12\textheight]{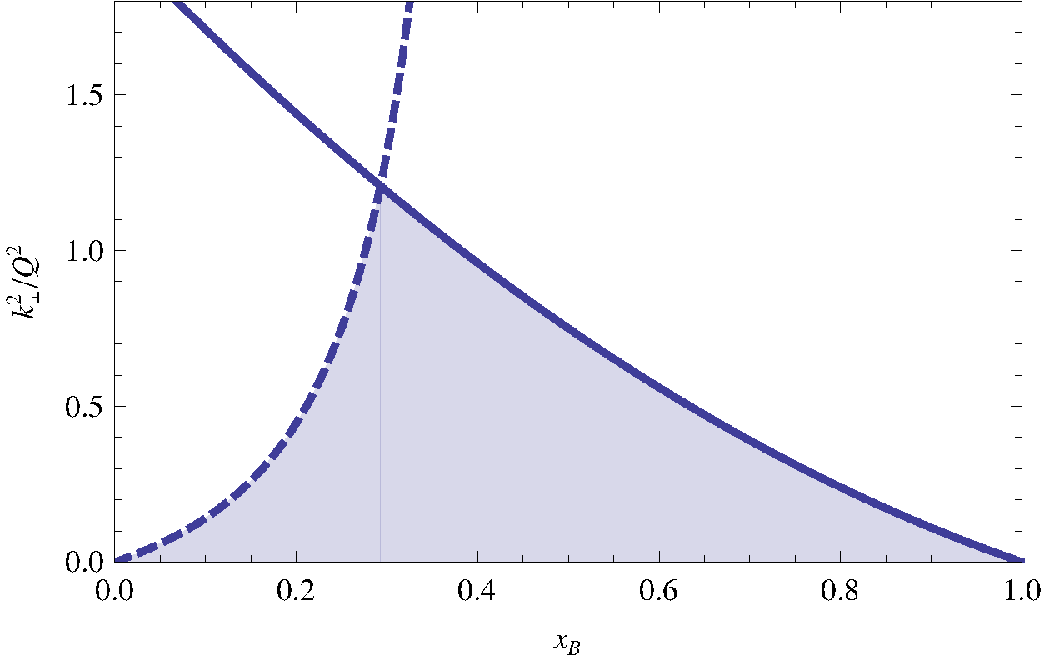}
\caption{
 $\kt^2/Q^2$ phase space as determined by the bounds
of Eqs.~\protect\eqref{cutenergy}
and \protect\eqref{cutdirection}. The allowed region, which fulfills  both bounds, is
 represented by the shaded area below the solid line.}
\label{cuts}
\end{figure}

The ratio $\kt^2/Q^2$, as constrained by Eqs.~(\ref{cutenergy})
and~(\ref{cutdirection}),
is shown in Fig.~\ref{cuts} as a function of $\xb$:
from this plot it is immediately evident that although in principle
Eq.~\eqref{cutdirection} (represented by the dashed line)
gives a stringent limit on $\kt^2/Q^2$ in the region $\xb<0.5$,
it intercepts the bound of Eq.~(\ref{cutenergy}) (solid line) in
$\xb\simeq0.3$,
where the latter becomes most relevant.
Notice also that present data from HERMES and COMPASS experiments span the
region $x_B \leq 0.3$,
where only the momentum bound of Eq.~\eqref{cutdirection} plays a role.

Once the maximum value of $k_\perp$ is limited by 
Eqs.~\eqref{cutenergy} and \eqref{cutdirection},
we set the appropriate normalization coefficient
\be
f_{q/p}(x,\kt) =  f_{q/p}(x) \, \frac{1}{1-e^{-(k_\perp^{\rm max})^2/\avk}}
 \, \frac{e^{-{\kt^2}/{\avk}}}{\pi \avk}\,,
\ee
where $(k_\perp^{\rm max})^2$ denotes the maximum value of $k^2_\perp$ for each
given values of $\xb$ and $Q^2$ as required by
Eqs.~\eqref{cutenergy}, and \eqref{cutdirection},
so that
\be
f_{q/p}(x) = \int_{0}^{2\pi} \!\! d \varphi  \int_{0}^{k_\perp^{\rm max}}
\!\!k_\perp \,d k_\perp \, f_{q/p}(x,\kt)\,.
\label{normalization1}
\ee

%
%
The average hadronic transverse momentum $\langle P_{T}^2\rangle$ of the final,
detected hadron $h$ is defined as:
\begin{equation}
\langle P_{T}^2\rangle=\frac{\int d^2 \mathbf{P}_T P_T^2 d\sigma }{\int d^2
\mathbf{P}_T d\sigma}\,\cdot\label{ptsmobs}
\end{equation}
Notice that if the integral 
is performed over the range $[0,\infty]$, then $\langle P_{T}^2\rangle$
coincides with the Gaussian width of the unpolarized $P_T$ distribution of
the SIDIS cross section: $\langle P_{T}^2\rangle\equiv \ptsmG \equiv \langle p_{\perp}^2\rangle+z^2_h\langle
k_{\perp}^2\rangle$.
The experimental $P_T$ range, however, usually span a finite region between some
$P_T^{min}$ and $P_T^{max}$; 
therefore, in any experimental analysis, one inevitably has
$\langle P_{T}^2\rangle \neq\ptsmG$,
even without considering the cuts in Eqs.~(\ref{cutenergy}) and
(\ref{cutdirection}). Consequently, the relation
$\langle P_{T}^2\rangle\simeq\langle p_{\perp}^2\rangle+z^2_h \langle
k_{\perp}^2\rangle$
holds only approximatively.

\begin{figure}[t]
\includegraphics[width=0.23\textwidth,angle=-90]{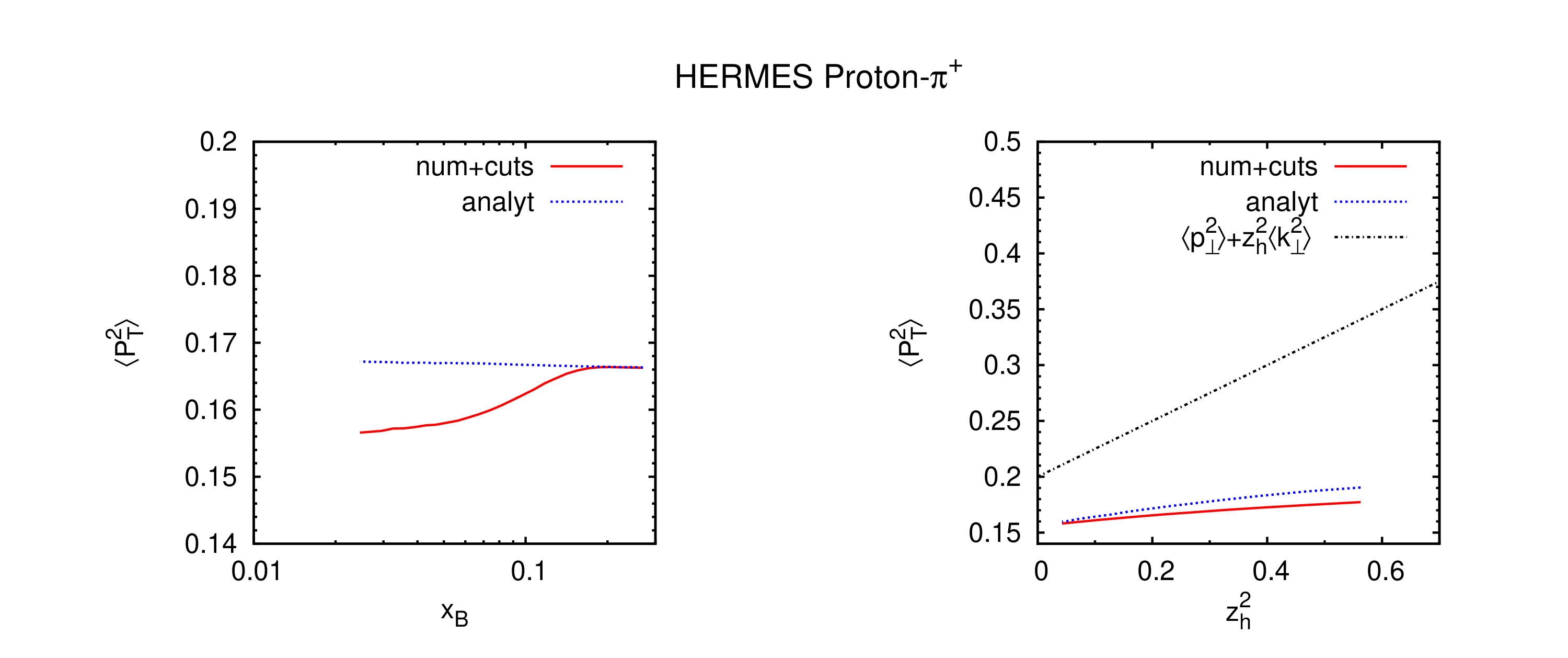}
\includegraphics[width=0.23\textwidth,angle=-90]{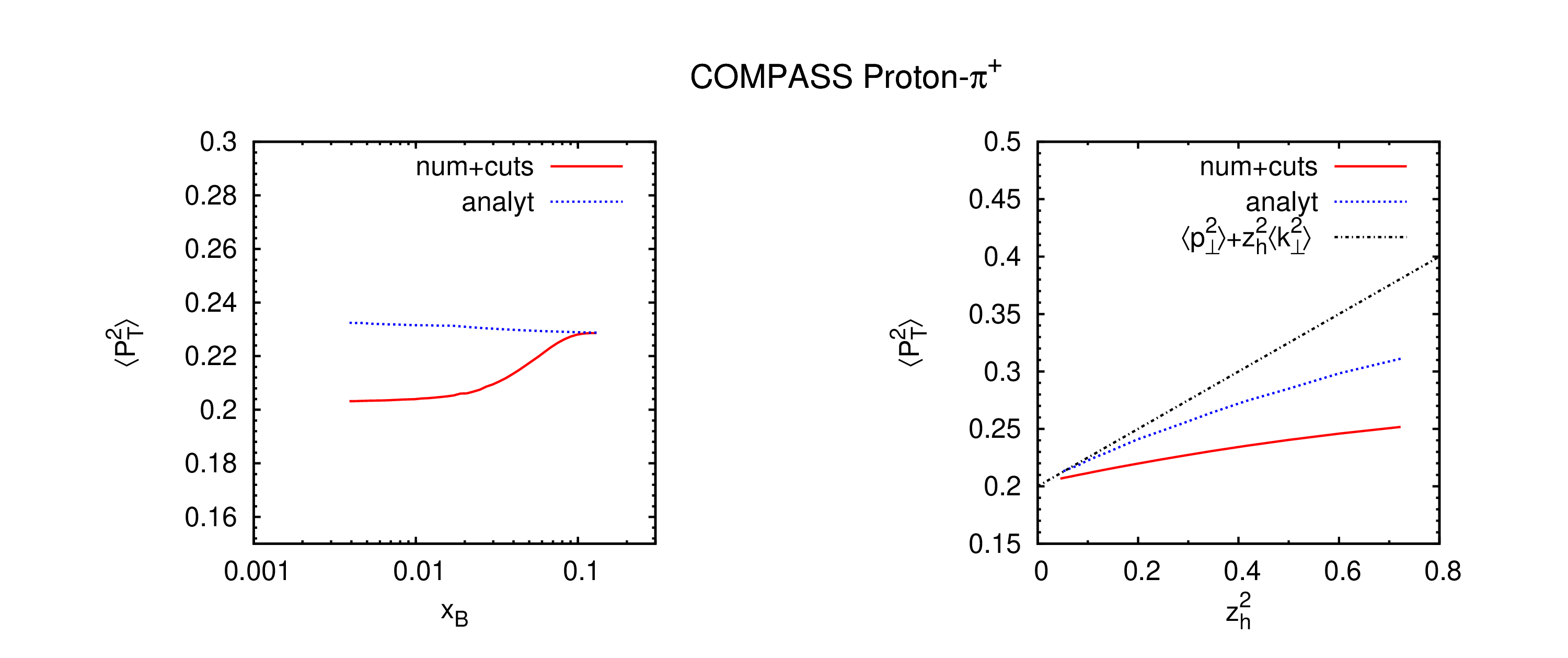}
\caption{$\langle P_{T}^2\rangle$
as a function of $\xb$ (left plot) 
and of $z_h^2$ (right plot), for $\pi^+$ production at HERMES (right panel) and
COMPASS (left panel). 
The solid (red) line corresponds to $\langle P_{T}^2\rangle$ calculated with a
numerical integration implementing Eqs.~(\ref{cutenergy}) and
(\ref{cutdirection}), while the dashed (blue) line is $\langle P_{T}^2\rangle$ 
calculated with an analytical integration. In both cases we have applied the 
corresponding experimental cuts. 
The dash-dotted (black) line corresponds to the Gaussian $\ptsmG$. \label{capt-x-ptsm}}
\end{figure}

Figure~\ref{capt-x-ptsm} shows the average hadronic transverse momentum $\langle
P_{T}^2\rangle$ as a function of $\xb$ and of $z_h^2$ for $\pi^+$ production 
at HERMES and COMPASS, respectively.
The  solid (red) lines correspond to $\langle P_{T}^2\rangle$ obtained
with a numerical $\kt$ integrations 
implementing Eqs.~(\ref{cutenergy}) and (\ref{cutdirection}). Instead, the
dashed (blue) lines correspond to $\langle P_{T}^2\rangle$ computed with an 
analytical integration. 
In both cases we have taken into account the appropriate experimental $P_T$ cuts.
Clearly, at low $\xb$, there is a substantial deviation from the analytical calculation,
which also affects the value of $\langle P_{T}^2\rangle$ as a function of
$z_h^2$. As far as the $z_h$ dependence is concerned, first of all, 
one can  see that there is a large deviation from the naive formula 
$\langle P_T^2\rangle=\langle p_{\perp}^2\rangle +z_h \langle k_{\perp}^2\rangle$, 
corresponding to the dash-dotted (black) lines, for both calculations. 
Secondly, although the $z_h^2$-dependence is not linear any more, it seems to be
approaching
an almost constant behavior 

The $\langle \cos \phi _h\rangle$ modulation receives two contributions,
both suppressed by one power of $(k_{\perp}/Q)$.
The Cahn term, which is proportional to the convolution of the unpolarized
distribution and fragmentation functions, was extensively studied in
Ref.~\cite{Anselmino:2005nn}.
There, EMC measurements~\cite{Ashman:1991cj} on the $\cos\phi _h$ modulation and
of the $P_T$ distribution on the unpolarized SIDIS cross section were used
to determine the Gaussian width of the $\kt$ distribution of the unpolarized
distribution function $f_{q/p}(x,\kt)$. 
The second term is proportional to the convolution of the Boer-Mulders
distribution function and the Collins fragmentation function and was neglected
in Ref.~\cite{Anselmino:2005nn}.
%
%

\begin{figure}[t]
\includegraphics[width=0.24\textwidth,angle=-90]
{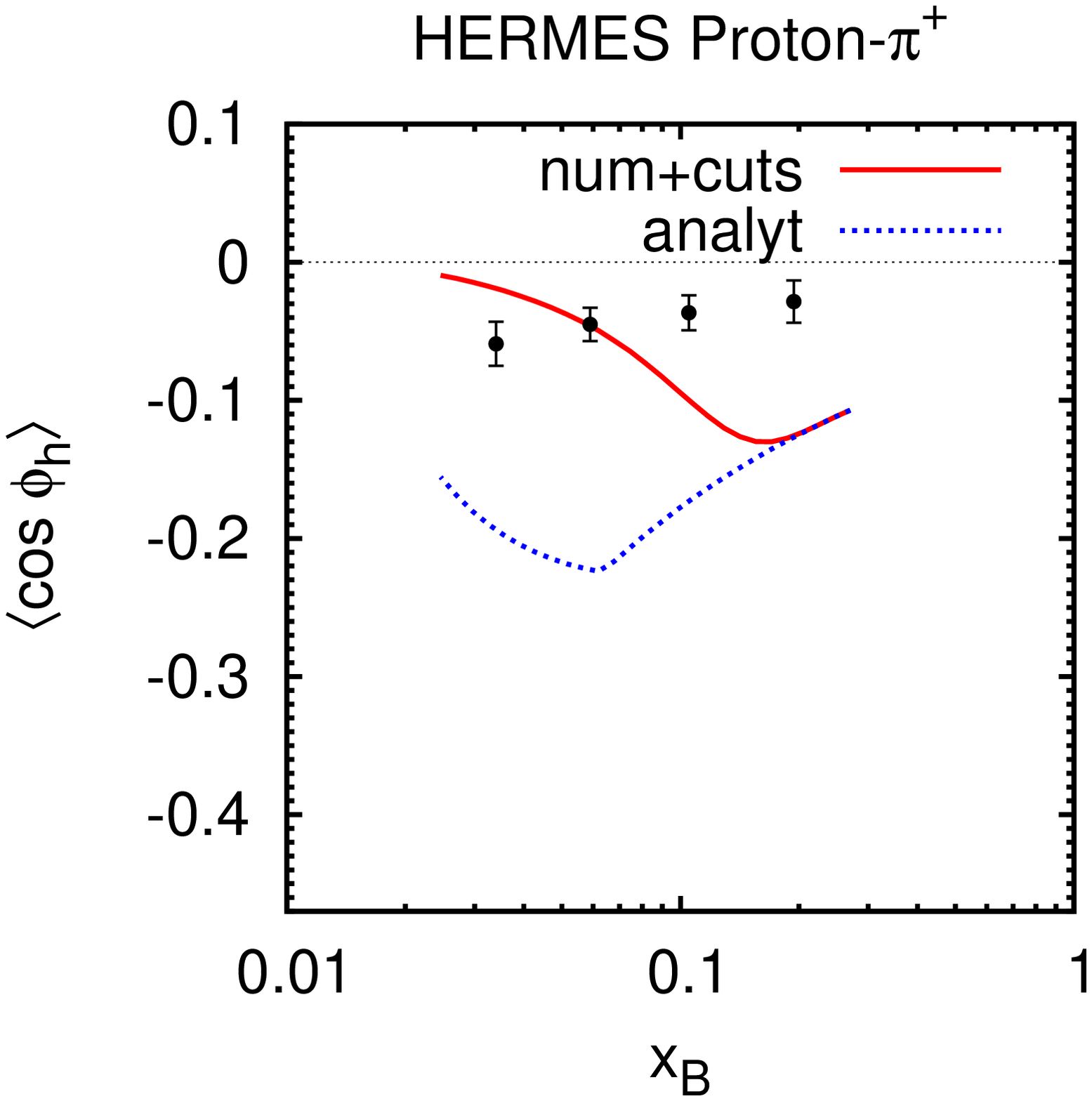}
\includegraphics[width=0.24\textwidth,angle=-90]
{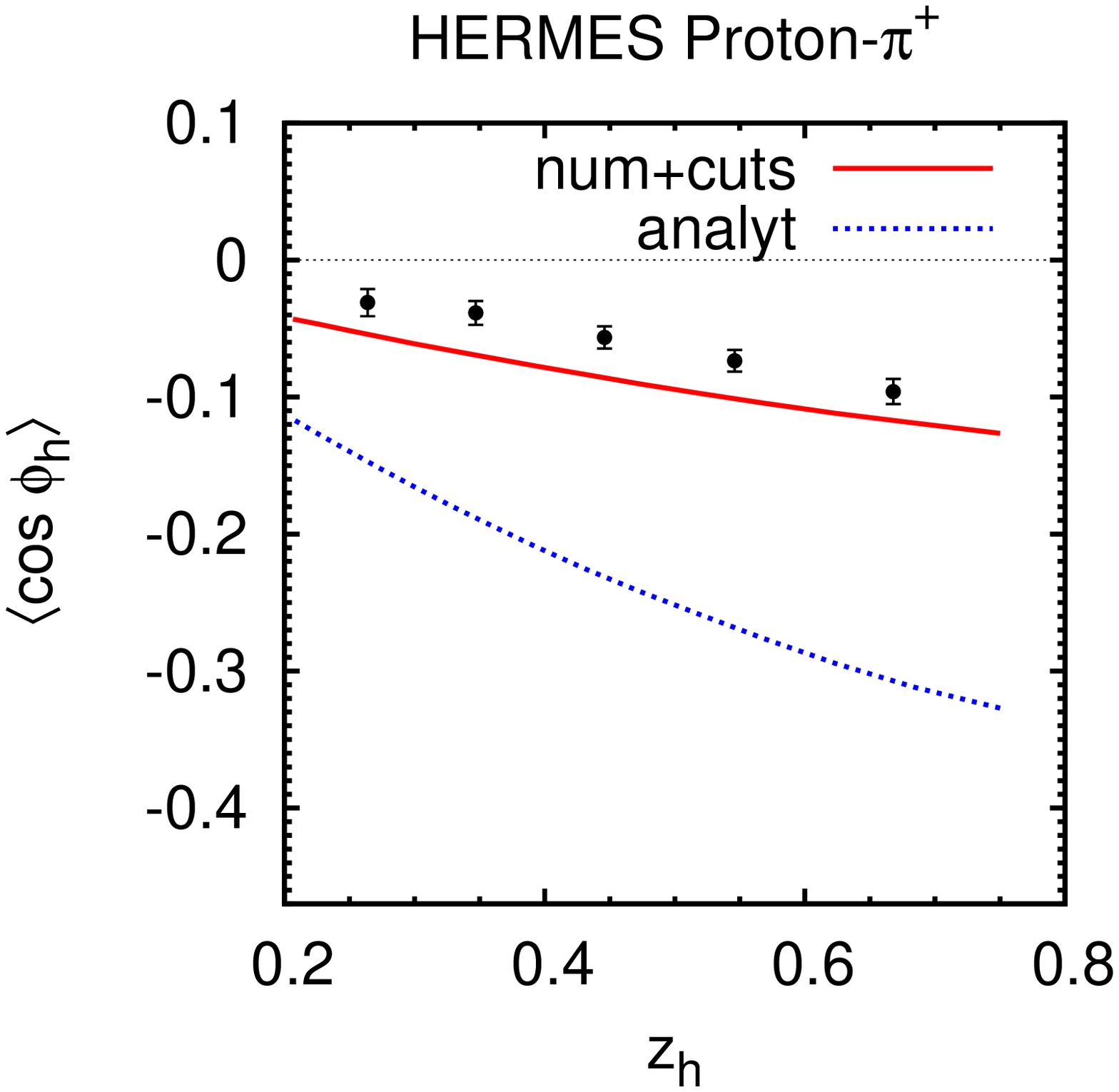}
\includegraphics[width=0.24\textwidth,angle=-90]
{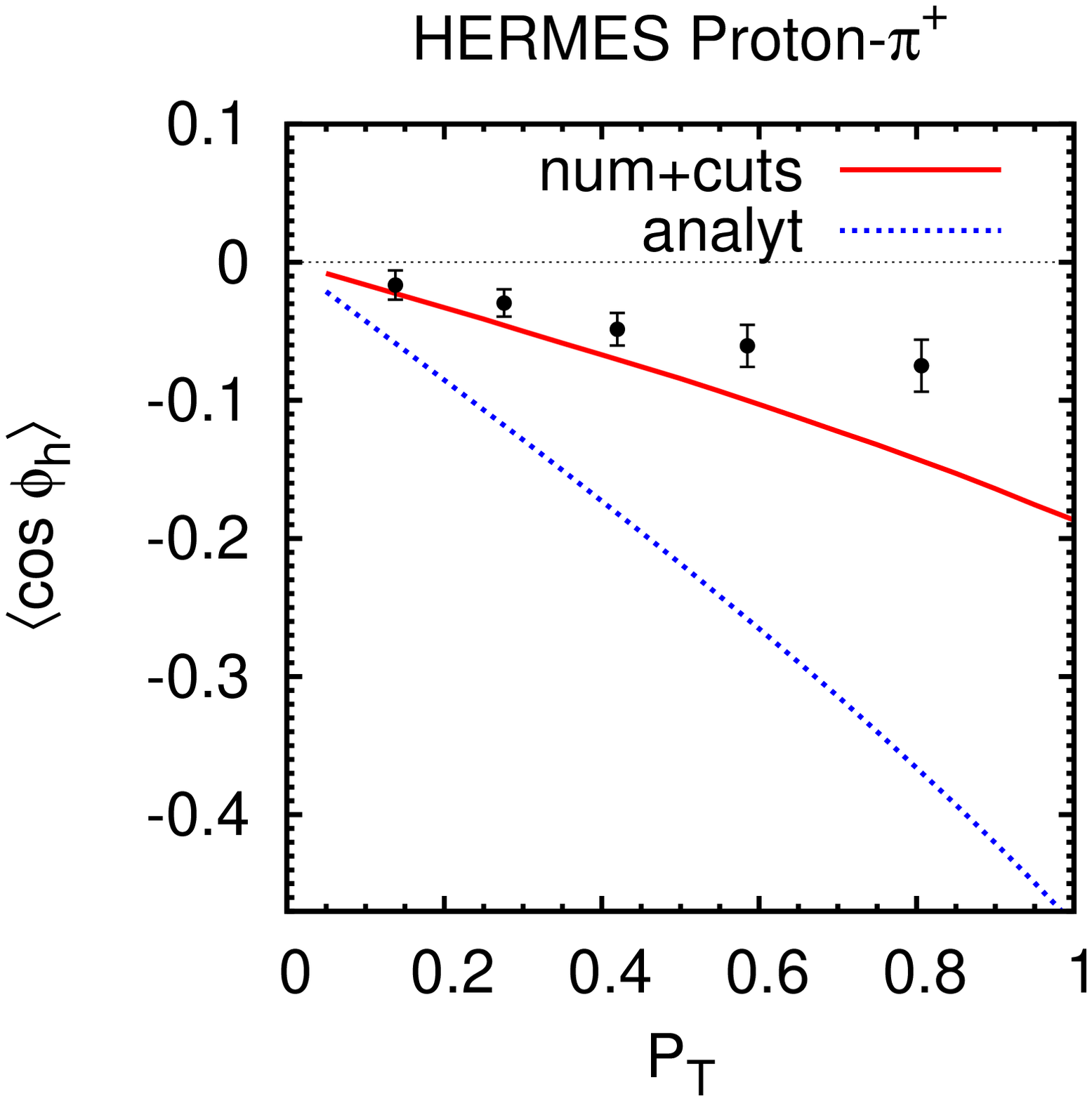}
\caption{\label{hermes-pip-cosphi}
Cahn contribution to $\langle\cos \phi _h\rangle$ for
$\pi ^+$ production at HERMES, 
as a
function of $\xb$ (left plot), $z_h$ (central plot) and $P_T$ (right plot).
The solid (red) line corresponds to $\langle \cos \phi _h\rangle$ calculated
with a numerical $k_\perp$ integration over
the range $[0,k_\perp^{max}]$.
The dashed (blue) line is $\langle\cos \phi _h\rangle $
obtained by integrating over $k_\perp$ analytically. 
The full circles are preliminary experimental data from
Ref.~\protect\cite{Giordano:2010zz}. }
\end{figure}

\begin{figure}[t]
\includegraphics[width=0.24\textwidth,angle=-90]
{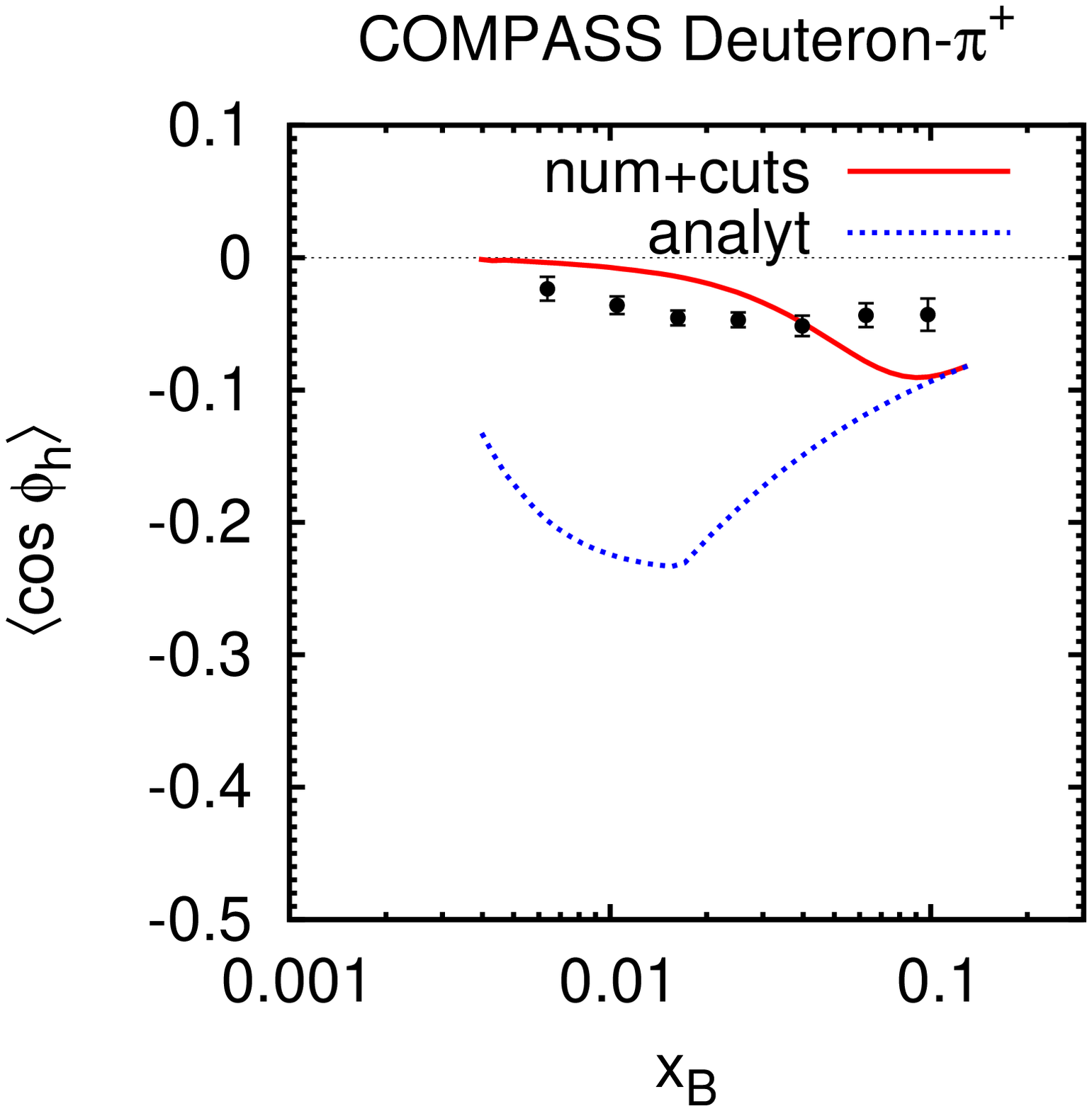}
\includegraphics[width=0.24\textwidth,angle=-90]
{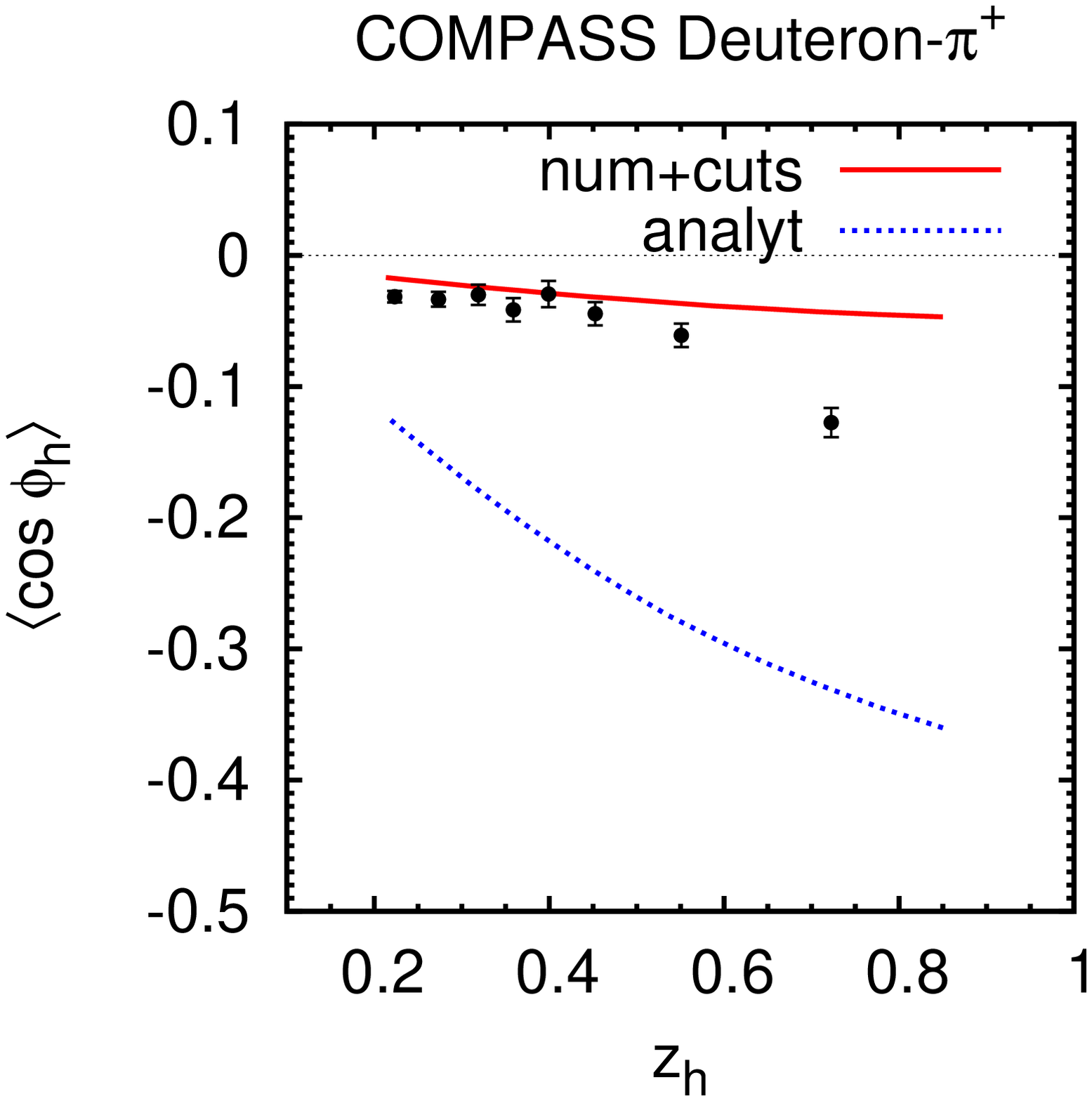}
\includegraphics[width=0.24\textwidth,angle=-90]
{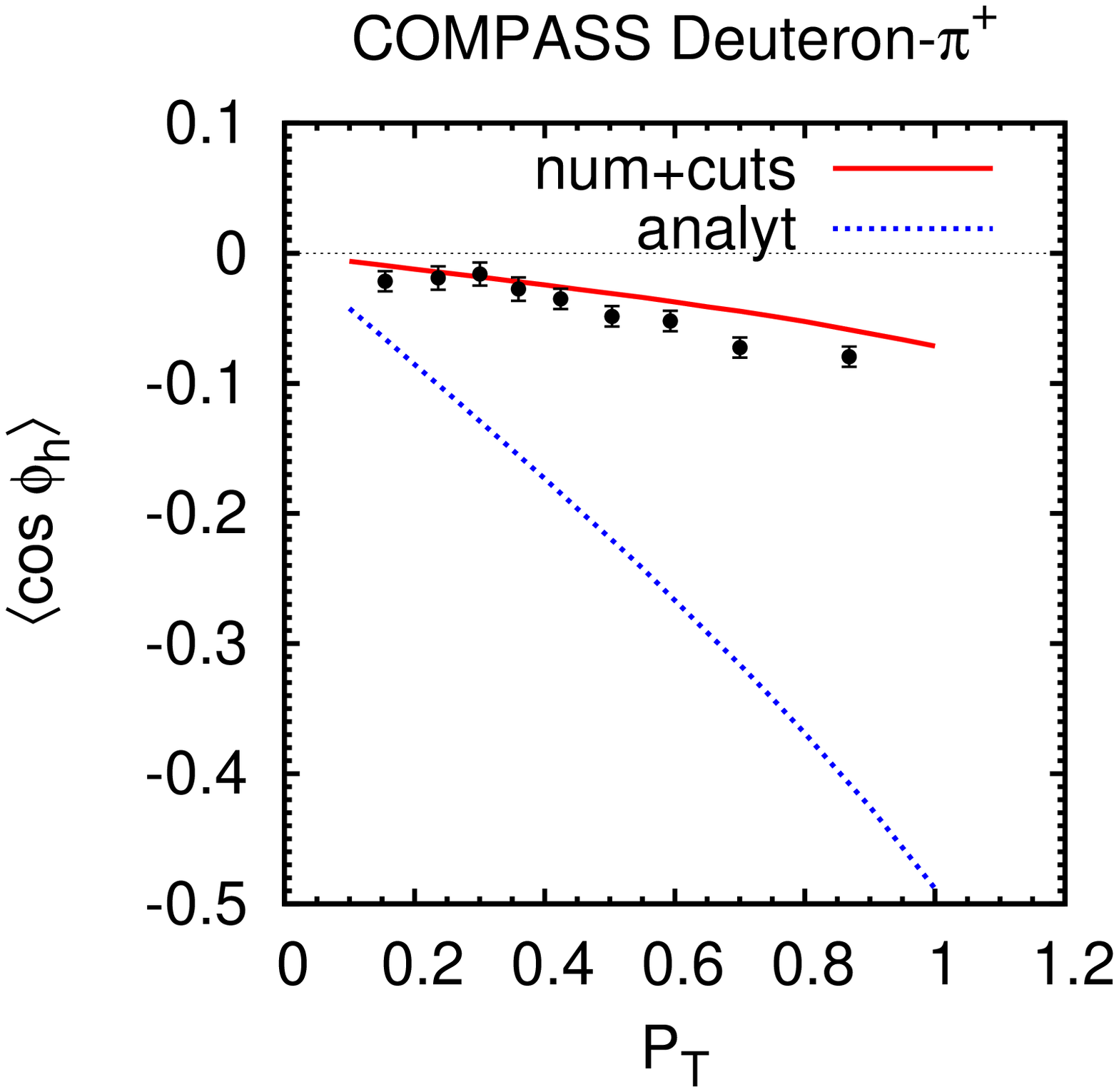}
\caption{\label{compass-pip-cosphi}
Same as Fig.~\protect\ref{hermes-pip-cosphi}, for the COMPASS kinematics.
The full circles are preliminary experimental data from Ref.~\protect\cite{Sbrizzai:2009fc}.}
\end{figure}

Figures~\ref{hermes-pip-cosphi} and~\ref{compass-pip-cosphi} 
show how a large deviation from the analytical
integration results
is obtained by applying the $\kt$ bounds of Eqs.~(\ref{cutenergy}) and
(\ref{cutdirection})
when computing the Cahn effect contribution to $\langle \cos \phi _h\rangle$
corresponding to the HERMES and COMPASS kinematics.
In these figures our results, obtained with and without $\kt$ -
cuts, are compared to the latest 
HERMES~\cite{Giordano:2010zz} and COMPASS~\cite{Sbrizzai:2009fc} data.
Although still showing a considerable deviation from the experimental data,
our calculation confirms that physical partonic cuts have a quite dramatic
effect in the small $x$ region,
and should therefore be taken into account in any further analysis of these
experimental data.

To evaluate the influence of the partonic cuts on the contribution to  $\langle
\cos \phi_h\rangle$ originating 
from the Boer-Mulders$\otimes$Collins term, 
we use the parametrization of Ref.~\cite{Anselmino:2008jk} for the Collins
function 
while for the Boer-Mulders function we apply the extraction of
Ref~\cite{Barone:2009hw}.
The Boer-Mulders contribution is very tiny 
and is not strongly affected 
by kinematical cuts of Eqs.~\eqref{cutenergy} and \eqref{cutdirection}.

The residual discrepancy between the model prediction and the measurements of
the $\langle \cos \phi_h\rangle$ azimuthal 
moment could indicate that higher twist contributions, from pure twist-3
functions, for example, might be non negligible in this modulation.
More elaborated phenomenological studies including twist-3 TMDs would be
necessary to confirm these observation.

%
%



\bibliographystyle{aipproc}   

\bibliography{newsample}

\end{document}